\documentclass{article}
\usepackage{graphicx}
\begin{document}
\def\bc{\begin{center}}
\def\ec{\end{center}}
\def\btab{\begin{tabular}}
\def\etab{\end{tabular}}
\def\mc{\multicolumn}
\def\ul{\underline}
\def\ol{\overline}
\def\bi{\bibitem}
\def\deg{$^{\circ}$}
\def\C{$^{\circ}$C}
\def\ra{\mbox{$\rightarrow$}}
\def\ef{\mbox{$E_{\rm F}$}}
\def\tc{\mbox{$T_{\rm c}$}}
\def\tmin{\mbox{$T_{\rm min}$}}
\def\tnot{\mbox{$T_{\circ}$}}
\def\etal{{\it et al}}
\def\ie{{\it i.e.,}}
\def\eg{{\it e.g.}}
\def\etc{{\it etc.}}
\def\vs{{\it vs}}
\def\wrt{{\it w.r.t.}}
\def\nbs2{NbS$_2$}
\def\2hnbs2{2H-NbS$_2$}
\def\3rnbs2{3R-Nb$_{1+x}$S$_2$}
\def\gaxnbs2{Ga$_x$NbS$_2$}
\title{Low temperature resistance minimum in non-superconducting \3rnbs2
and 3R-\gaxnbs2}
\author{Asad Niazi\protect$^{1}$\ and A. K. Rastogi\protect$^{2}$\\
{\it School of Physical Sciences, Jawaharlal Nehru University,} \\
{\it New Delhi - 110067, India}}
\date{}
\maketitle
\footnotetext[1]{Presently Visiting Fellow, Dept. of Condensed Matter
Physics and Material Science, T.I.F.R., Mumbai, India,
Email: asad@tifr.res.in.}%
\footnotetext[2]{Corresponding author, Email: rastogi@jnuniv.ernet.in,
Fax: +91-11-6194137.}
\begin{abstract}
We report the structural and electron transport properties of \3rnbs2
($x \geq .07$) and 3R-\gaxnbs2 ($.1 \leq x \leq .33$) prepared as
polycrystalline pellets as well as single crystals grown by vapour
transport. We observe a resistance minimum in these compounds between
20--60~K, with the \tmin\ proportional to $x$. The resistance scales as
$\rho/\rho_{\rm min}(T/\tmin)$ between $.2 < T/\tmin < 2$ for
different phases with $x \leq .25$ whose resistivity differs by an
order of magnitude. Powder X-ray diffraction (XRD) also shows
progressively increasing intensity of superlattice lines with cation
concentration. The thermopower changes sign around the resistance
minimum. The explanation of the resistance minimum and the simultaneous
rapid suppression of superconductivity is sought in $e$-$e$ scattering
effects in the presence of cation disorder in these narrow band
anisotropic materials.
\end{abstract}

71.45.Lr -- Charge density wave systems.\\
74.62.Bf -- Effects of material synthesis, crystal structure, and
chemical composition.\\

\noindent{\bf Running Head:} Low temperature resistance minimum, \etc
\newpage

\section{Introduction}
The layered transition metal dichalcogenides
(LTMDs) MX$_2$ of the Group V metals (M = V, Nb, Ta; X = S, Se) and
their intercalation compounds have been the subject of numerous
studies on the inter-relationship between superconductivity and charge
density waves (CDW), both of which arise from the strong
electron-phonon ($e$-$ph$) coupling within the layers
\cite{wilson1,friend}. Parameters such as stoichiometry, polymorphism,
disorder and intercalation have been extensively used to study the
physical properties of these low dimensional compounds. There is
however, no satisfactory explanation for their effect on the above
transitions.

Amongst the binary compounds, all di-Selenides (V, Nb, Ta) and all
polymorphs of TaS$_2$ show CDW transitions, while the 2H and 4H Nb and
Ta compounds are also superconducting. \2hnbs2 and 1T-VS$_2$ are
unusual -- In the former, any CDW is suppressed below the
superconducting $\tc \sim$ 6.2~K due to $e$-$e$ interactions in the
narrow unhybridised Nb $d$ bands. 1T-VS$_2$ is structurally metastable,
supposedly due to reduced covalency which destabilizes the layered
structure in its stoichiometric composition. The effects of $e$-$e$
interactions are also significant in 1T-VSe$_2$ in which the normal to
incommensurate CDW transition temperature \tnot\ increases from 110~K
under pressure induced broadening of the narrow V $d$ bands. It also
shows a small Curie-like contribution in $\chi$ from a small excess of
V between the layers. 2H-NbSe$_2$ ($\tnot \sim$ 33~K), 2H-TaS(Se)$_2$
($\tnot \sim$ 75(122)~K) and 4H$_{\rm b}$-TaS(Se)$_2$ ($\tnot \sim$
22(75)~K) also exhibit $\tc \sim$ 7.2~K, $\leq$~.6~K and $\sim$ 1~K
respectively. High pressure lowers \tnot\ and raises \tc\ towards its
`undistorted' values (\cite{friend} and references therein).
Lattice-dynamics calculations suggest that in these materials,
$e$-$ph$ renormalisation effects on phonon frequencies are crucial in
causing lattice instability, as well as raising \tc\ \cite{nishio1}.
The above transitions are absent in di-Tellurides in which strong
intermetallic bonding and metal atom clustering leads to more stable
but distorted structures.

Stoichiometry, intercalation and disorder significantly affect the CDW
and superconductivity in these compounds. In 1T-TaS$_2$ a mere 50~ppm
of isoelectronic Nb impurities were observed to destroy the long range
phase coherence of the commensurate CDW and completely suppress the
associated metal-insulator (MI) transition \cite{zwick}. On
intercalation, the \tc\ in Ta compounds increases from $<$ 1~K upto 5~K
while it is lowered for Nb \cite{revel}--\cite{gamble}. In
self-intercalated 2H-Nb$_{1.05}$Se$_2$ the \tc\ was suppressed below
2.2~K \cite{revel}, while 2H-NbSe$_2$(EDA)$_{1/4}$ showed no
superconductivity but a resistance minimum at $\sim$ 25~K
\cite{meyer}. Post-transition metal intercalated \2hnbs2 also did not
show superconductivity \cite{karnezos}. `Kondolike' resistance minima
at 20~K along with a CDW were observed for 2H-Fe$_{.05}$Nb(Ta)Se$_2$
\cite{dai}. Li$_x$NbS$_2$ ($0 \leq x \leq .5$) has a complex
dependence of \tc\ on $x$ due to interplay of polymorphic changes
(2H--3R) and electron transfer effects upon intercalation
\cite{mcewen}. In fact, comparable disorder induced resistance minima
at low temperatures are also seen in structurally similar Graphite
intercalation compounds, and explained using weak localisation and
$e$-$e$ interactions \cite{piraux}.

A less studied polymorph of \nbs2 is the 3R phase (space group $R3m$).
While the inherently stoichiometric \2hnbs2 forms at high temperatures
($\geq 950$\C) under high S vapour pressure, any excess Nb (at low S
vapour pressure) results in the metal-rich 3R phase \cite{fisher}. The
stoichiometry limit for single phase \3rnbs2 depends on the
preparation temperature, and a minimum of $x = .03$ has been reported
at 650\C\ \cite{tatsuki}. Samples prepared at high temperature and
reported closer to stoichiometry are essentially a 2H--3R mixture with
properties correspondingly in between. The 3R phase can thus be
considered a self-intercalated phase with the excess Nb in octahedral
interlayer vacancy sites \cite{powell}. While no superconductivity has
been observed in \3rnbs2 down to 1.7~K, there is also no evidence of a
CDW which could depress the same. Conduction in LTMDs is essentially
intralayer. Hence the role of even small amounts of interlayer metal
in suppressing superconductivity between different polymorphs needs to
be investigated to understand better the effects of polymorphism,
stoichiometry, disorder and dimensionality on the properties of these
compounds.

We have prepared \2hnbs2, \3rnbs2 ($x \simeq .07$) and 3R-\gaxnbs2 ($x
= .1, .25, .33$). \2hnbs2 has \tc\ $\sim$ 6.2~K, as expected. The
metal-rich \3rnbs2 shows no superconductivity; rather, a resistance
minimum $\sim$ 20~K. Non-magnetic Ga-intercalated 3R phases exhibit
similar low temperature resistance minima (20~K $\leq \tmin <$ 60~K).
Thermopower also shows anomalies around \tmin. We observe a clear
correspondence of \tmin\ and extent of localisation to the
cation-to-anion ratio and the preparation temperature, \ie\ to the
extent of structural distortion from the ideal layered host.

\section{Experimental Details}
\subsection{Preparation and characterisation}
The compounds were prepared by heating stoichiometric mixtures of the
pure elements (Ga 99.999\%, Nb 99.7\%, S 99.999\% pure, CERAC) in
evacuated quartz ampoules. \2hnbs2 was obtained using 10\% excess
Sulfur. Chemical analysis by ICP-AES and EDX techniques gave Fe $<$
240~ppm, Mn $<$ 15~ppm and Cr $<$ 10~ppm. Pressed polycrystalline
pellets were sintered at different temperatures -- 950\C\ for the
binary compound, and 850\C\ (LT phase) \& 1100\C\ (HT phase) for
Ga-intercalated compounds -- and quenched to room temperature. Single
crystal flakes were obtained by vapour transport (950 \ra\ 900\C) for
\2hnbs2 and 3R-Ga$_{.1}$NbS$_2$. Some Sulfur rejection was observed at
high temperature in all cases, indicating a metal rich composition.

\begin{figure}[htbp]
\centering
\includegraphics[width=15cm]{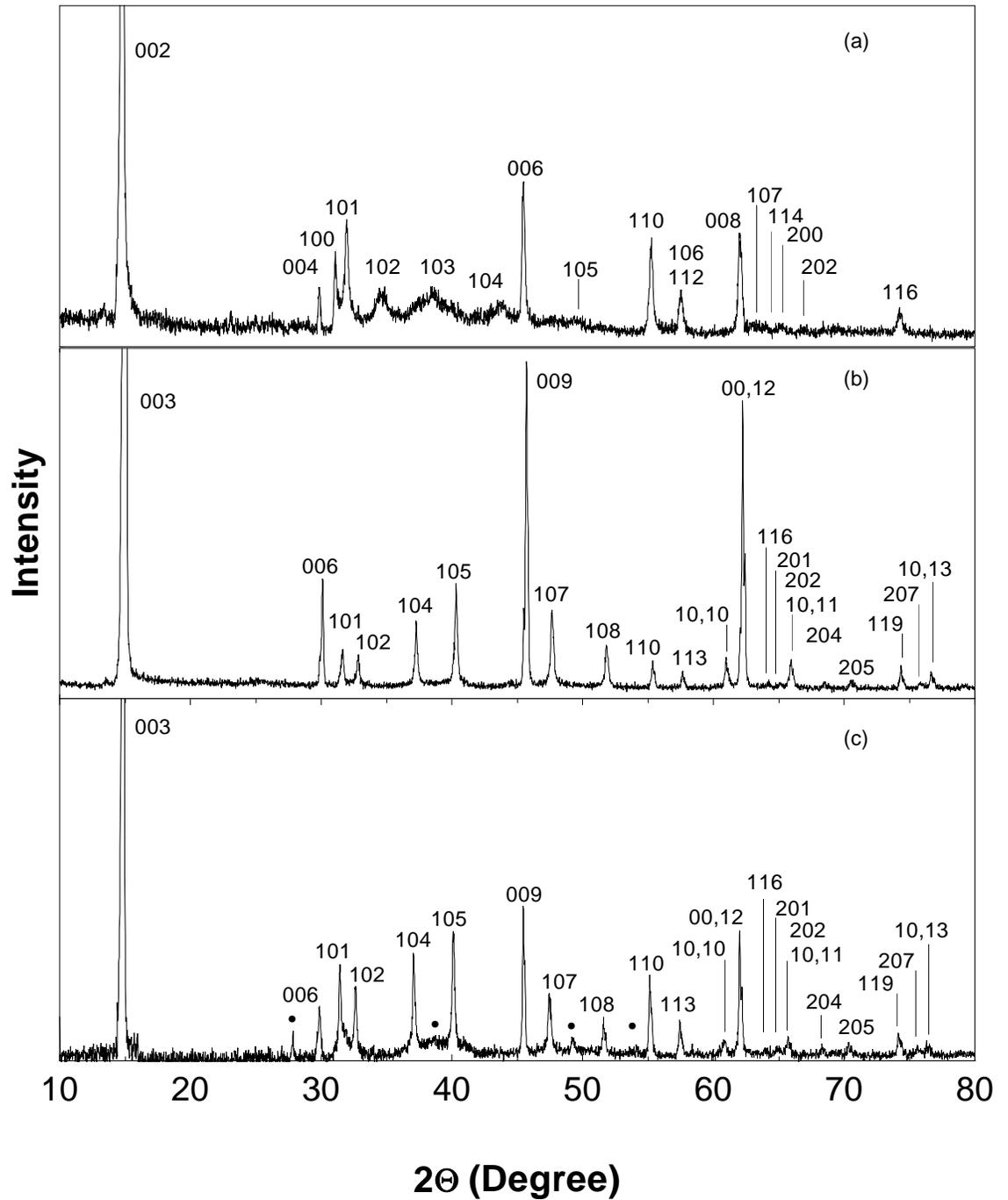}
\caption{Powder X-ray patterns of (a) \2hnbs2, (b) \3rnbs2 and
(c)3R-Ga$_{.1}$\nbs2 (LT). The (002) and (003) peaks were very strong
due to preferred orientation and are truncated to magnify the rest. The
broad peaks of the 2H phase, (a), are in contrast to the sharp lines
of the metal-rich 3R phases, (b) and (c), as discussed in the text.
Emerging
superlattice lines ($\bullet$) can be seen in (c).}
\label{fig1}
\end{figure}

The compounds were characterised by room temperature powder XRD. We
compare the XRD patterns of \2hnbs2, \3rnbs2 and 3R-Ga$_{.1}$\nbs2(LT)
in Fig.~\ref{fig1}. The 2H phase ($a$ = 3.32~\AA, $c$ = 11.97~\AA) had
broad ($10l$) lines indicating well known disorder in the stacking of
the weakly coupled antiparallel layers \cite{jellinek}. The metal rich
3R phases on the other hand had sharp lines, the HT phases more ordered
than the LT ones. The excess metal (Nb\,/\,Ga) occupying the
octahedral interlayer vacancies \cite{karnezos,powell} pins the layers
together and prevents stacking faults. \3rnbs2 ($a$ = 3.32~\AA, $c$ =
17.88 \AA) and 3R-\gaxnbs2 ($x$ = .1, .25, .33(LT phase)) ($a$ =
3.33$_5$~\AA, $c$ = 17.90$_5$~\AA) have similar lattice parameters.
However, the Ga-intercalated phases exhibit superlattice reflections
proportional in number and intensity to Ga content and the preparation
temperature, indicating progressive distortion. The .33~Ga (HT) phase
($a$ = 7.19~\AA, $c$ = 17.30~\AA) is structurally very different, with
about 10\% reduction in volume and a doubling of the $a$-axis. The
superlattice lines order into a doubled $a$-axis and show much lower
$c/a \simeq 1.603$. Thus with increasing cation concentration and
preparation temperature, the trigonal prismatic close-packed layers
transform towards a distorted octahedral coordination of Nb by
staggered Sulfur layers. Progressive metal clustering and vacancy
formation within the Nb layers finally leads to a well ordered phase
with a different structure. This is similar to cation-rich
Nb$_2$Se$_3$, Nb$_3$S$_4$, Cu$_{.33}$\nbs2, \etc\ which have strong
Nb-Nb bonds giving rise to zig-zag chains\,/\,clusters. These changes
are also reflected in the electronic properties. The exact nature of
the Nb atom clustering in our compounds would require a more detailed
structural study.

\subsection{Electronic properties}
The electronic properties were studied by 4-probe d.c.-resistance
(4.2--300~K), thermopower ($S(T)$) (14--300~K) and magnetic
susceptibility ($\chi(T)$)(80--300~K) measurements. The
d.c.-resistance of pellets as well as single crystal flakes was
measured in van der Pauw geometry \cite{vdpauw} using Ag-paste
contacts. The $S(T)$ of pellets pressed between Cu stubs was measured
in the differential mode. Absolute $S$ was determined by calibrating
with Pb \cite{roberts}, and correcting for the Cu leads. The $\chi(T)$
was measured on compacted powders under a 9.7~kOe field in a
vibrating sample magnetometer (VSM) \cite{vsm}. The results are
summarised in Table~\ref{tbl1}.

The samples showing resistance minimum obviously do not
follow Mattheissen's rule for scattering from dilute impurities under
the quasi-free electron approximation. Therefore, we consider the
$\rho_{\rm min}$ as characteristic of the residual resistance, and in
subsequent discussions the residual resistance ratio (RRR) is defined
by $\rho_{300}/\rho_{\rm min}$; while, for the superconducting samples
the RRR = $\rho_{300}/\rho_{\tc}$, where $\rho_{\tc}$ is the value of
the resistivity at \tc.

\begin{figure}[htbp]
\centering
\includegraphics[width=15cm]{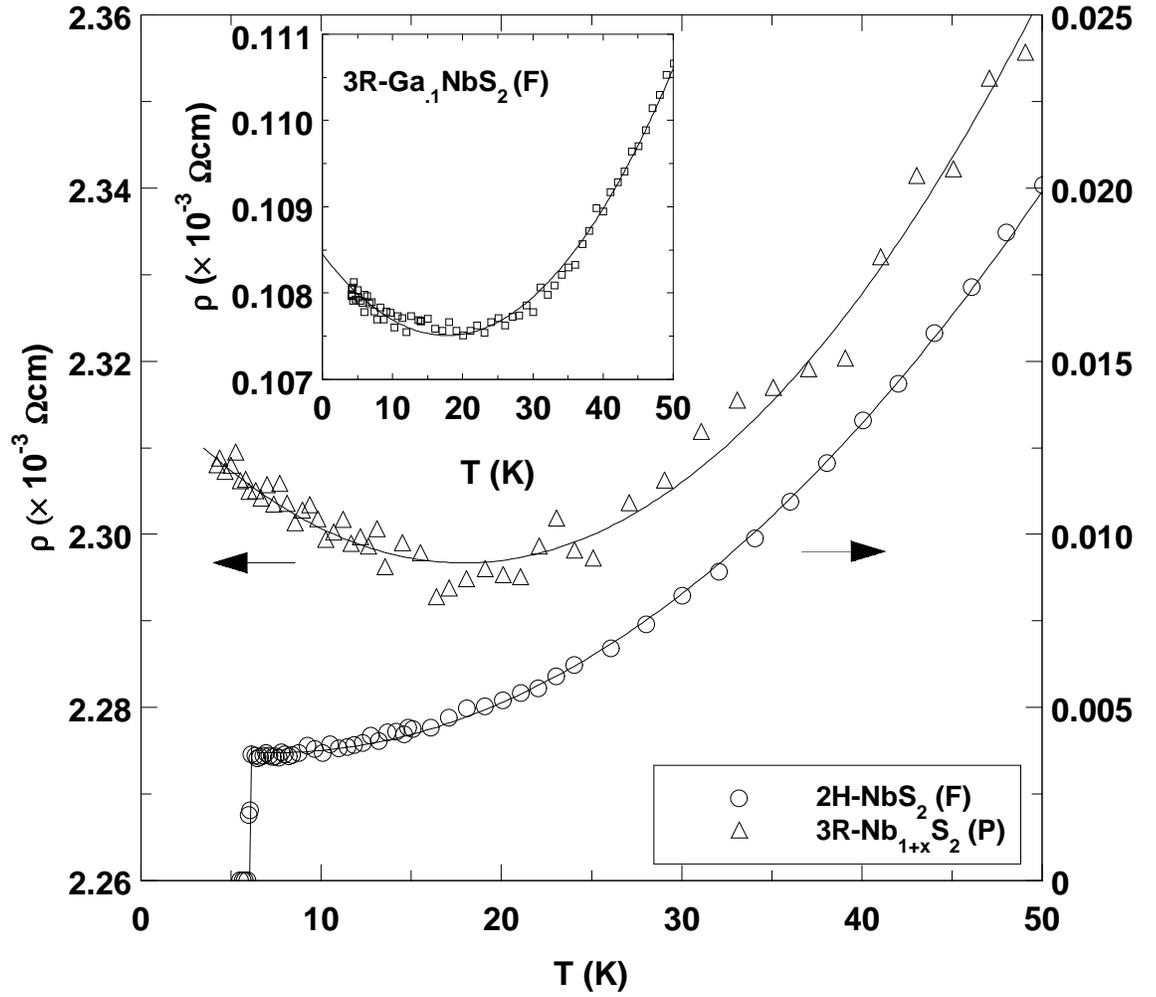}
\caption{Resistivity in \2hnbs2
single crystal flake (F), \3rnbs2 pellet (P) and (inset)
3R-Ga$_{.1}$\nbs2 flake. The 3R phases exhibit $\rho_{min}$ at $\sim$
20~K. The solid lines are a guide to the eye.}
\label{fig2}
\end{figure}

\begin{figure}[htbp]
\centering
\includegraphics[width=15cm]{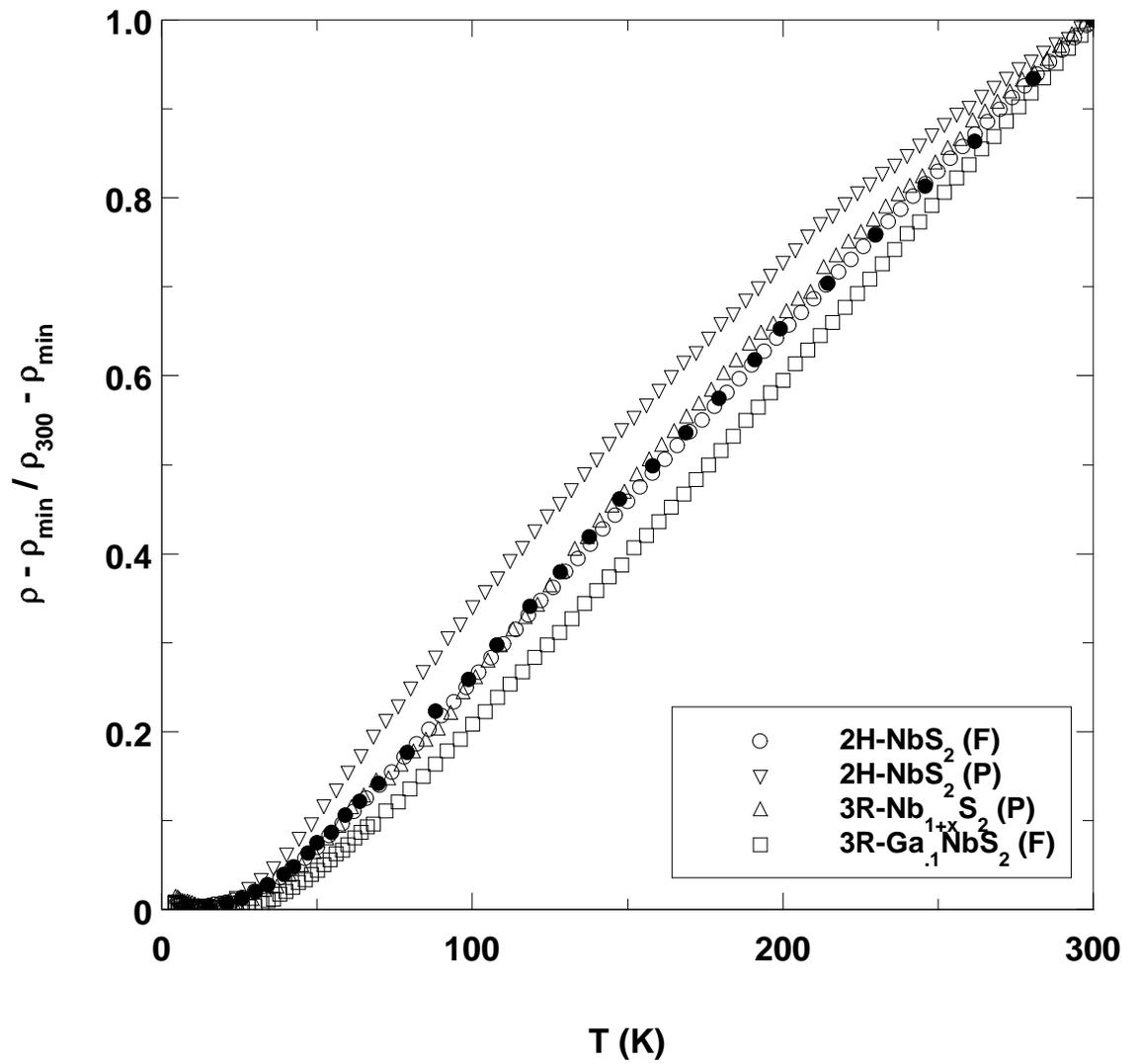}
\caption{Normalised `ideal' Resivitivity vs Temperature in the pellets
and flakes of the binary 2H and 3R phases and the .1 Ga
flake along with data ($\bullet$) from \protect\cite{naito}.}
\label{fig3}
\end{figure}

The resistivity results on single crystal flakes and polycrystalline
pellets of both 2H and 3R polymorphs are shown in Fig.~\ref{fig2} and
\ref{fig3}. Grain boundary scattering in the pellets
and the anisotropy of conduction in flakes would prevent their
inter-comparison. Figure~\ref{fig3} therefore shows the normalised
resistance behaviour after subtracting low temperature residual
resistances. The 2H phase \tc\ $\simeq 6.2$~K and its large RRR of
$\sim 69$ for in-plane conduction
in the flakes compare well with earlier studies \cite{naito,hamaue}.
Thus inspite of a considerable degree of stacking disorder, the
essentially in-plane conduction and also the superconductivity is not
affected in 2H polymorphs. The 3R phases on the other hand have no
stacking disorder as inferred from their sharp XRD pattern, but gave a
resistance minimum around and above 20~K. The temperature and extent
of the minimum increases with the amount of intercalate atoms.

The overall conduction behaviour of the 3R phases showing a resistance
minimum is also significantly different. Firstly, we notice a large
increase in residual resistance upon intercalation (Table~\ref{tbl1}).
For example, the .1~Ga crystal flake measured along the plane has a RRR
of only 1.65, although its resistivity is actually much lower than
superconducting  \2hnbs2 (RRR $\sim$ 69). The scattering of electrons
at higher temperatures is also significantly different. We observe in
the 3R phases a comparatively faster drop in resistance of flakes as
well as pellets and also a larger contribution of T$^2$ - term above
their resistance minimum (Fig.~\ref{fig3}). The reduction in saturation
effects in the high temperature conduction of the polycrystalline
\3rnbs2 phase probably indicates an increase in its isotropy of
conduction. A proper explanation of this difference would require a
detailed investigation of the effects of non-stoichiometry and
polymorphic changes on the band structure and $e$-$ph$ interaction
effects. The observed differences, however, cannot be simply related to
doping of carriers since extra Nb or Ga atoms have similar effects. We
presently conclude that in the intrinsically non-stoichiometric
\3rnbs2 phases the intralayer scattering potentials are significantly
increased, leading to large residual resistance, increased $e$-$e$
scattering effects, and consequently low temperature localisation of
carriers.

\begin{figure}[htbp]
\centering
\includegraphics[width=15cm]{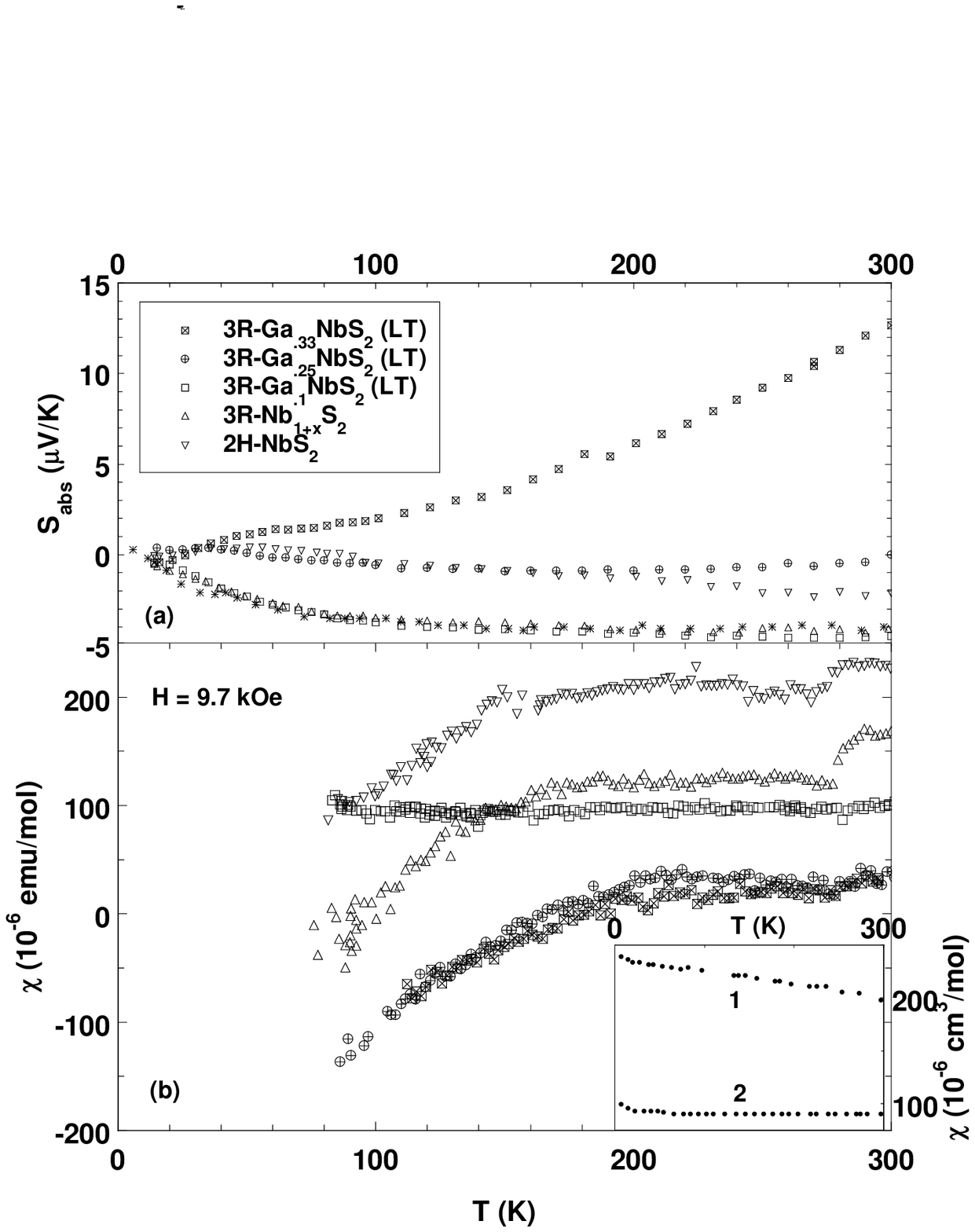}
\caption{(a). $S(T)$ and (b). $\chi(T)$, in \2hnbs2, \3rnbs2 and
3R-\gaxnbs2 (LT phases), $x = .1, .25, .33$. In (a) data from
\protect\cite{bowmeester}, ($\ast$), compares well with our \3rnbs2 and
.1~Ga phases. In (b) the inset shows data from \protect\cite{fisher},
(1) \2hnbs2 and (2) 3R-Nb$_{1.07}$S$_2$. The diamagnetic contribution
of core-electrons is expected to be $\sim 100 \times 10^{-6}$~emu/mole.}
\label{fig4}
\end{figure}

The $S(T)$ and $\chi(T)$ results on polycrystalline 2H and 3R phases
are shown in Fig.~\ref{fig4}. The $S(T)$ of 3R phases of self
intercalated and .1~Ga intercalated phase are very similar but differ
markedly from that of \2hnbs2. For the former a large, fairly
constant, negative $S(T)$ at high temperature changing rapidly towards
positive below 100~K can be seen. For $\chi(T)$ we have subtracted a
small saturated contribution which was observed in the $M(H)$
behaviour at low fields ($< 1$~kOe). Our data differ from previous
studies, as we observe a continuous reduction in $\chi(T)$ instead of
a slight increase on cooling \cite{fisher}. These are the typical
dependences observed in compounds showing CDW instabilities.

\section{Discussion}
We have earlier mentioned that group V TMD compounds of other than the
present studies, \ie\ layered 1T-VSe$_2$, 2H-NbSe$_2$ and various
polymorphs of TaS(Se)$_2$, show varying degrees of CDW formation. The
associated anomalies in their transport and magnetic properties are
much weaker than in 1D CDW structures such as NbS$_3$, TaS$_3$ etc.
For example, except for 1T-TaS$_2$ which has a CDW lead MI transition,
the usual increase in resistance at \tnot, is barely observable and
only a steep increase in its slope is observed below \tnot. Recent
optical studies on 2H-TaSe$_2$ confirm the absence of any abrupt
formation of a charge-excitation gap at \tnot\ \cite{vescoli}. The
steep increase in the resistance slope below \tnot\ is found to be a
consequence of freezing-out of scattering channels since the Drude
scattering peak in $\sigma(\omega)$ ($\omega$ \ra\ 0) becomes narrower
below \tnot. In the light of the above observations, we cannot rule
out CDW-correlations in our 3R phases, more so since superstructural
distortions are clearly observed in intercalated phases. The
thermopower and magnetic susceptibility variation also suggest some
non-magnetic electronic correlations developing on cooling. A careful
structural study of the 3R polymorphs of \nbs2 at low temperature is
therefore required to ascertain the relation of the superlattice with
the observed transport and magnetic properties.

\begin{figure}[htbp]
\centering
\includegraphics[width=15cm]{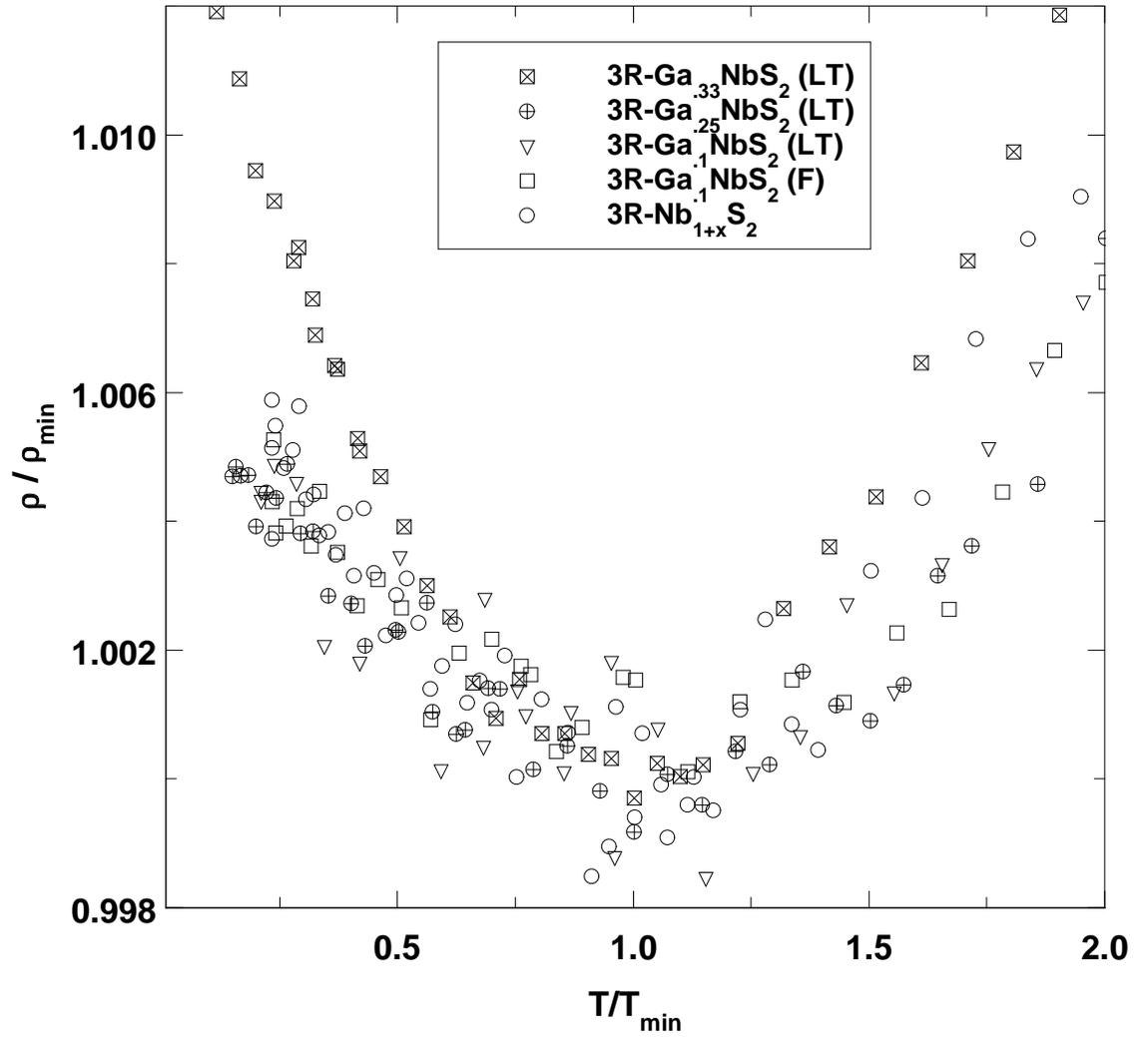}
\caption{Normalised Plot of Resistivity vs Temperature for \3rnbs2 and
3R-\gaxnbs2 (LT), $x = .1, .25, .33$. There is close scaling below
\tmin\ except for the metal rich .33~Ga phase.}
\label{fig5}
\end{figure}

In Fig.~\ref{fig5} we show a scaled plot of resistance behaviour with
temperature for the compounds of our study between $.2 < T/\tmin < 2$.
There is seen to be a close similarity for the crystal flakes and
pellets of compounds having different intercalate (Ga/Nb)
concentration and showing vastly different resistivity values. It
clearly indicates the role of defects in these compounds since
$\rho_{\rm min}$ and \tmin\ show a systematic increase with the
intercalate concentration.

The resistance minimum may be sought to be explained by Kondo
scattering from dilute magnetic impurities such as Fe or Mn
\cite{kondo,heeger,maple}. However, as stated earlier, the maximum
concentration of such impurities in our compounds ($< 250$~ppm) is too
low for significant spin-flip scattering contribution. Neither do we
observe the expected logarithmic dependence of resistivity upon
temperature for $T < T_{min}$. The temperature of the Kondo minimum
plotted as $\rho/\rho_{\rm min}$ vs $T$ is in fact only weakly
dependent upon the extrinsic impurity concentration. Instead, we find
a strong dependence of \tmin\ on intercalate concentration which
scales as shown in Fig.~\ref{fig5}. Moreover, our $S(T)$ is smoothly
varying in the temperature range of resistance minimum instead of the
broad, large maxima expected in Kondo systems. The Pauli like
$\chi(T)$ of our compounds at high temperature is also in contrast to
the Curie-Weiss like behaviour of Kondo systems for $T \gg T_{\rm
min}$. Thus Kondo scattering cannot be invoked to explain the observed
behaviour of our compounds.

On the other hand, disorder and Coulomb interaction effects on the
scattering of electrons in a narrow anisotropic band of these
compounds are expected to be quite important. For $k_{\rm F}l \le 1$,
($k_{\rm F}$ -- Fermi momentum, $l$ -- elastic mean free path), the
quasi-classical treatment of elastic scattering leading to
Mattheissen's rule in dilute alloys breaks down. The rise in
resistance on cooling can be understood in terms of either increasing
interference corrections for the elastically scattered waves from the
static disorder or subtle changes in the excitation spectrum at the
Fermi level caused by $e$-$e$ interaction effects in the presence of
ionic disorder, as first suggested by Altshuler and Aronov
\cite{altshuler}. The relative importance of these quantum
correlations and their temperature dependence depends on
dimensionality and the details of band structure of the conduction
electrons.

\begin{figure}[htbp]
\centering
\includegraphics[width=15cm]{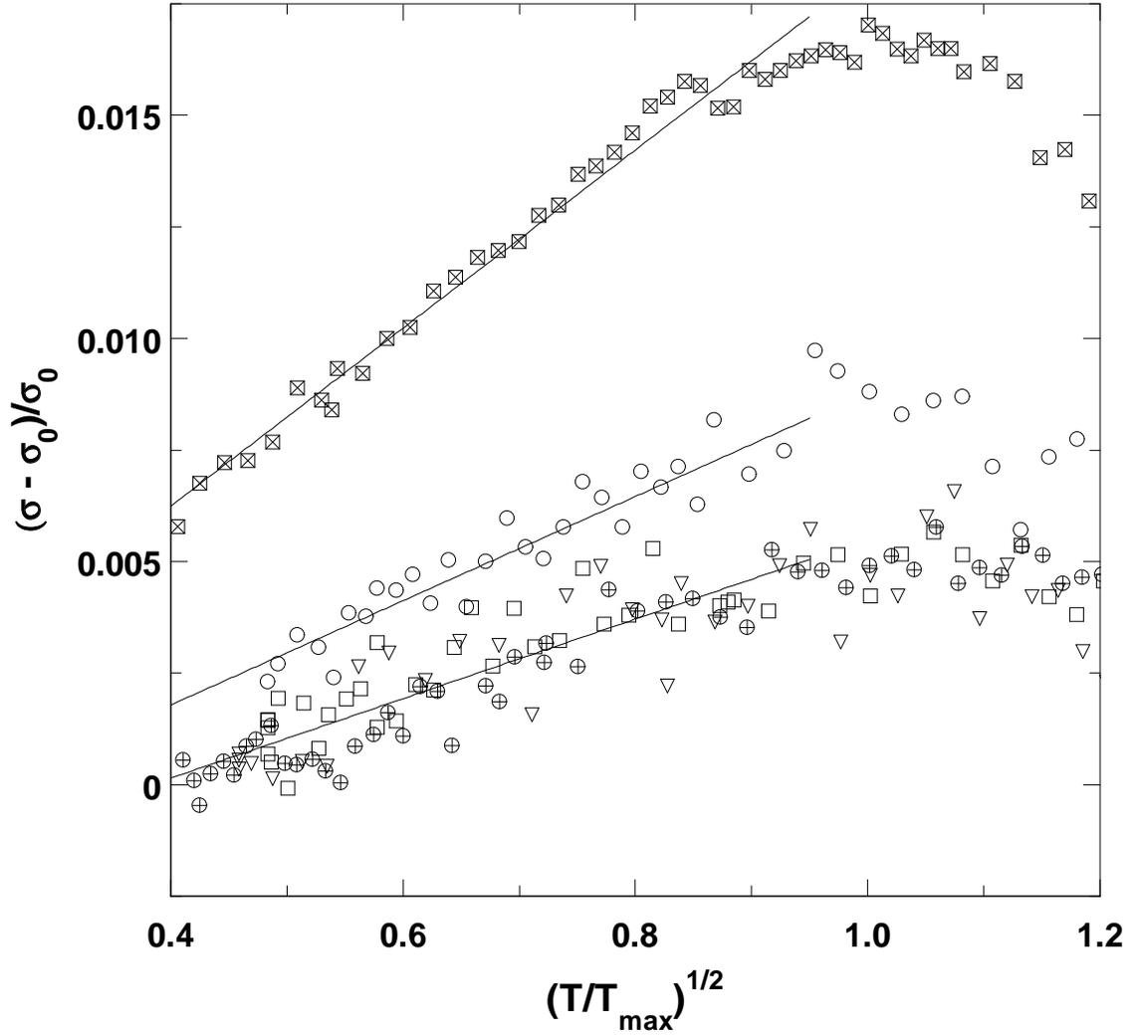}
\caption{Normalised Plot of Conductivity vs $(T/T_{\rm max})^{1/2}$ for the above phases. The figure legend is same as Fig.~\ref{fig5}. The solid
lines are guides to the eye. All curves scale well as $T^{1/2}$ below
$T = T_{\rm max}$. The slope of the curves depends on the stoichiometry 
details discussed in the text.}
\label{fig6}
\end{figure}

The plots of $\Delta \sigma / \sigma_{0}$ at low temperature are shown
in Fig.~\ref{fig6}, where $\sigma_{0}$ is the value of conductivity at
0~K obtained by extrapolating below 4.2~K. Our preliminary results show
a nearly $T^{1/2}$ dependence of the conductivity far below its maximum,
the coefficient of the $T^{1/2}$ term depending upon the stoichiometry
and structural details. The observed behaviour suggests quantum
corrections to the DOS due to long range Coulomb interactions between
conduction electrons \cite{altshuler}. The metal rich .33~Ga (HT) phase
also exhibits similar behaviour, though with a much larger slope. Here
the localisation effects are dramatically increased due to the clustered
nature of the metallic lattice.

We mention here an interesting possibility which may increase the long
range Coulomb interaction effects between charge carriers. The theory
of Altshuler and Aronov gives a depression in DOS(\ef)
\cite{altshuler}. The present compounds are prone to CDW correlations
which also reduce DOS(\ef). However, for CDW the $e$-$e$ interactions
are very special since they require $e$-$ph$ coupling across a nesting
wave vector. Therefore, the role of incipient CDW fluctuations in the
presence of disorder to give the observed resistance minimum behaviour
in these compounds should be seriously explored.

To summarise, we have observed low temperature resistance minima in
non-superconducting metal-rich \nbs2 derivatives -- non-stoichiometric
\3rnbs2 (\tmin $\sim$ 20~K) as well as 3R-\gaxnbs2 (20~K $\leq \tmin <$
60~K). A common physical origin of these minima is evident in the scaling
of resistance as $\rho/\rho_{\rm min}(T/\tmin)$ between $.2 < T/\tmin < 2$
for different phases whose resistivity differs by an order of magnitude.
The low temperature behaviour of these non-magnetic compounds cannot be 
explained by Kondo scattering effects. Instead, we find that the
conductivity varies as $T^{1/2}$ below its maximum (\ie\ below
\tmin($\rho_{\rm min}$)). We, therefore, propose that a possible cause for
the observed behaviour in these narrow band anisotropic systems is a
correction to the DOS(\ef) due to $e$-$e$ interaction effects in the
presence of ionic disorder. 

{\bf Acknowledgements:} We are grateful to Prof. Deepak Kumar for
illuminating discussions. AN thanks CSIR, New Delhi for financial
support during his doctoral work.

\newpage

\begin{table}[tbp]

\bc
\btab{|l|llll|l|l|}%
\hline %
\mc{1}{|c|}{\it Sample} & \mc{4}{c|}{\it D.C. Resistivity} &
\mc{1}{c|}{$S_{300}$} & \mc{1}{c|}{$\chi_{300}$}  \\
\cline{2-5}& $\rho_{300}$ & $\rho_{min}$ & T$_{min}$ &
R.R.R. & ($\mu$V\,/\,K) & ($\times 10^{-6}$ \\
& \mc{2}{l}{($\times 10^{-3}~\Omega$-cm)} & (K) &
($\frac{\rho_{300}}{\rho_{min}}$) & & emu/mole) \\
\hline
2H-NbS$_2$ ${\ddag}$ & 0.241 & 0.0035 ${\dag}$ & --  & 68.83& -- & -- \\
2H-NbS$_2$ & 2.02 & 0.80 ${\dag}$ & -- & 2.52 & $-$2.092 & $\simeq$ 200 \\
3R-Nb$_{1+x}$S$_2$ & 3.23 & 2.296 & 20 & 1.408 & $-$4.114 & $\simeq$ 100 \\
Ga$_{.1}$NbS$_2$ ${\ddag}$ & 0.178 & 0.1075 & 20 & 1.656 & -- & --\\
Ga$_{.1}$NbS$_2$ (LT) & 1.96 & 1.380 & 20 & 1.420 & $-$4.501 & $\simeq$ 100 \\
Ga$_{.25}$NbS$_2$ (LT) & 2.29 & 1.874 & 28 & 1.222 & $-$0.199 & $\simeq$ 30 \\
Ga$_{.25}$NbS$_2$ (HT) & 4.32 & 3.575 & 32 & 1.207 & 1.305 & $\simeq$ 35 \\
Ga$_{.33}$NbS$_2$ (LT) & 4.74 & 3.950 & 41 & 1.200 & 12.667 &$\simeq$ 30 \\
Ga$_{.33}$NbS$_2$ (HT) & 12.00 & 10.60 & 58 & 1.128 & 3.012 & $\simeq$ 140 \\
\hline %
\etab

\ec
\caption{Summary of Electronic Transport Properties of \gaxnbs2.
HT\,/\,LT : High\,/\,Low Temperature prepared phases. $\ddag$: Single
crystal. $\dag$: For superconducting \2hnbs2, R.R.R. =
$\rho_{300}/\rho_{\tc}$.}
\label{tbl1}
\end{table}


\end{document}